\documentclass[sigconf]{acmart}

\copyrightyear{2026}
\acmYear{2026}
\setcopyright{cc}
\setcctype{by-nc-nd}
\acmConference[]{}{}{}

\acmDOI{}
\acmISBN{}

\usepackage{array}
\usepackage{algorithmic}
\usepackage{graphicx}
\usepackage{textcomp}
\usepackage{xspace}
\usepackage{multirow}

\usepackage{pifont} 

\newcommand{\greybox}[1]{%
 \par\addvspace{0.5em}
  \noindent
  {%
    \setlength{\fboxsep}{2pt}
    \color{gray!70}%
    \fbox{%
      \colorbox{gray!10}{%
        \parbox{0.96\linewidth}{%
        \small
          \color{black}
          #1
        }%
      }%
    }%
  }%
    \par\addvspace{0.5em}
}

\newcommand{\etal}{\emph{et~al.}\xspace}

\def\BibTeX{{\rm B\kern-.05em{\sc i\kern-.025em b}\kern-.08em
    T\kern-.1667em\lower.7ex\hbox{E}\kern-.125emX}}

\begin{document}

\title{Fairness-First Design Thinking for Software Architecture}

\author{Iffat Fatima}
\affiliation{%
  \institution{Vrije Universiteit Amsterdam}
  \city{Amsterdam}
  \country{The Netherlands}}
\email{i.fatima@vu.nl}

\author{Markus Funke}
\affiliation{%
  \institution{Vrije Universiteit Amsterdam}
  \city{Amsterdam}
  \country{The Netherlands}}
\email{m.t.funke@vu.nl}

\author{Patricia Lago}
\affiliation{%
  \institution{Vrije Universiteit Amsterdam}
  \city{Amsterdam}
  \country{The Netherlands}}
\email{p.lago@vu.nl}

\renewcommand{\shortauthors}{I. Fatima, M. Funke, P. Lago}

\begin{abstract}
Fairness issues often remain hidden in digital systems, making them difficult to detect and even more difficult to address. In this study, we introduce a fairness-first Design Thinking (DT) approach to support addressing fairness concerns in software architecture (SA) design. We implemented our approach in a graduate-level course where students executed all steps of our DT approach as part of an assignment. We analyzed the assignment data to reflect on the implications for applying the DT approach in SA and teaching the DT approach in SA education. As a result of this study, we provide (i) a DT approach for SA, (ii) implications of the DT approach on handling fairness in both problem and solution spaces, and (iii) implications for education. Our reflections highlight that fairness theory and context identification are essential for a holistic, fairness-first design. We propose the use of composite views to address cross-cutting concerns such as fairness. In the future, we will update the course material to provide end-to-end fairness traceability in SA, helping students to understand how fairness concerns can be translated into actionable design decisions. 
\end{abstract}


\maketitle
\section{Introduction}
Digitalization has reshaped the way that fairness issues manifest in a digital society. The root causes of these issues are often hidden by the complexity and opacity of digital systems~\cite{Abrams2025}. As a result, we see an increase in the prominence of fairness concerns in different sectors, where technology not only overlooks existing issues but can also reinforce them~\cite{Hurlin2024-creditscoring}. Such an amplification of unfairness can lead to negative systemic impacts in society at large. For example, in healthcare, diagnostic systems have been shown to exhibit racial bias, leading to disparities in health outcomes that lead to long-term allocation of unequal health benefits~\cite{Obermeyer2019-bias-health}. In the financial sector, credit scoring algorithms have been shown to embed historical biases that disproportionately affect people in marginalized communities~\cite{Hurlin2024-creditscoring}. In the criminal justice sector, predictive algorithms have been criticized for perpetuating overpolicing in communities of color or with a migration background~\cite{jefferson2020digitize}. In the employment sector, recruiting algorithms trained on historical recruitment data have been found to reinforce gender imbalance and racial inequalities in the workforce~\cite{Kelly-Lyth2021}. Content recommendation systems optimized for engagement have been criticized for leading to situations in which vulnerable users are exposed to content that contradicts their intent and negatively affects their well-being~\cite{Golbeck2025}.
Furthermore, the unchecked development and use of AI systems raise broader ethical questions related to privacy, transparency, environmental sustainability, and many more. 

These examples are only a glimpse of the broader challenges of fairness in digitalized sectors. Although legal frameworks such as the AI Act and GDPR provide accountability mechanisms for digital systems, it is equally crucial to embed fairness as a foundational principle in software design~\cite{deSouzaSantos2025--biassmells}~\cite{Alidoosti2022}. Proactive alignment of digital systems with the fairness needs of society would ensure that fairness is not treated merely as a regulatory requirement but as a fairness-first commitment integrated into the software design process. Research shows that fairness-first design thinking (DT) is necessary to reduce the fairness debt in digital systems, which may have long-range effects~\cite{deSouzaSantos2025--biassmells}. Once the software is implemented and operationalized, the changes required to solve the fairness issues go beyond the technical aspects of the software to the social, governance, and legal domains~\cite{deSouzaSantos2025--biassmells}, adding further complications. 

The issues mentioned above create the need for processes to support a fairness-first DT to address fairness issues during software design. As a first step in this direction, we introduce a DT approach for fairness-first DT in software architecture (SA). We applied our DT approach in an educational setting where graduate-level students used it to explore fairness issues and formulate solutions as part of a course assignment. The study presents the DT approach, the results of its application, and our reflections on SA and education. 

\section{Background}
\label{sec:bg}
This section provides some background on fairness, DT, and the role of fairness in the design of digital systems.

\subsection{\textbf{Fairness in the Theory of Philosophy}}
Fairness is a fundamental concept in philosophy, social justice, and ethics that has evolved throughout history. The early mentions of fairness emerge from the distributive justice concepts of Aristotle, where all individuals are treated proportionally according to their merit or need, rather than strictly equally~\cite{aristotle2014nicomachean}. Later theoretical developments expanded the theory of fairness into multiple dimensions. Rawls~\cite{rawls1971theory} formalized distributive justice as a principle to allocate resources fairly across society, while procedural justice theorists such as Leventhal~\cite{leventhal1980procedural} emphasized fairness in the processes used for decision making and resource allocation. Bies and Moag~\cite{bies1986interactional} introduced interactional fairness, focusing on the quality of interpersonal treatment. Greenberg~\cite{greenberg1993justice} later extended this to informational fairness, highlighting the role of transparency and explanation in fairness judgments. Retributive fairness, on the other hand, concerns the fairness of punishments and corrective actions~\cite{carlsmith2002psychological}. Together, these frameworks offer a comprehensive lens for evaluating fairness in outcomes, processes, communication, and interpersonal dynamics, especially relevant to the design of sociotechnical systems.
These theories suggest that fairness means treating people without favoritism or prejudice, but it can also involve differentiating treatment based on important factors such as need, effort, or merit to ensure outcomes that are just and balanced. Hence, context is extremely important in the application of fairness principles. 

\subsection{\textbf{Design Thinking: Origins and Ethics Dimension}}
The concept of DT first appeared in 1969 by Herbert Simon~\cite{simon1969}. DT is a methodology to create solutions with a human-centered approach~\cite{plattner2010designthinking}. Its interdisciplinary character makes it suitable for answering complex \textit{`wicked problems'} such as ethics and fairness ~\cite{koh2015designthinking}. Ethics is a core component of human-centered design because it involves engaging with people, especially the intended users or beneficiaries~\cite{steen2013codesign}. Hence, their moral values and experiences must be taken into account during the design process. DT emphasizes understanding stakeholder concerns, their context, redefining problems, and creating innovative solutions that fulfill stakeholder needs. It has five stages ~\cite{dSchool2009}: (i) \textit{emphathize}, i.e., to understand the needs of people in the context of the design challenge, (ii) \textit{define}, i.e., capturing the design problem, (iii) \textit{ideate}, i.e., to synthesize ideas that help solve the problem, (iv) \textit{prototype} i.e., an artifact that showing a solution, and (v) \textit{test}, i.e., to ensure the prototype serves the intended purpose of solving the problem.

\subsection{\textbf{Fairness in Digital Systems}}

In digital systems, fairness is mostly studied for algorithmic decision-making. Researchers have developed fairness metrics to formally assess and compare fairness in digital systems. Common metrics include (i) group fairness metrics, which assess whether different demographic groups are treated equally, and (ii) individual fairness metrics, which assess whether similar individuals receive similar outcomes~\cite{dwork2012fairness}. These fairness metrics are sometimes conflicting and may not be suitable for all contexts \cite{kleinberg2016inherent}. As digital systems become more integrated into decision making, the choice and interpretation of fairness metrics become increasingly critical~\cite{corbett2018measure}.

\subsection{Fairness from a Software Architecture Lens}
In this section, we give an overview of concepts from SA theory and literature that help build the foundations of our study. 

Software architecture is defined as a \textit{set of design decisions}~\cite{Jansen2005}. These decisions are often represented as architectural diagrams (or design views), where each view represents a certain architectural aspect. In SA, viewpoints serve as a structured template or modeling conventions to represent stakeholder concerns and associated guidelines for a particular kind of architectural representation. A design view is a concrete realization of the viewpoint as a model or diagram that illustrates a certain architectural aspect through the lens established by the viewpoint~\cite{rozanski2012}. These design views help capture architecturally significant quality concerns.

Similar to sustainability~\cite{Lago2015-sus-quality}, fairness can be seen as an architectural quality concern~\cite{Vora2025} which is (i) \textit{multidimensional}~\cite{santos2025softwarefairnessdebt} encompassing both social and technical aspects and (ii) \textit{cross-cutting}~\cite{rozanski2012} affecting multiple architectural elements and design views. To explore these multidimensional and cross-cutting concerns, such as fairness, the Sustainability Assessment Framework Toolkit (SAF) \cite{Lago2024-SAF-Toolkit} offers a set of thinking tools. For instance, Decision Maps (DMaps) are a graphical representation of the system features and their associated quality concerns across different dimensions (economic, environmental, social, and technical). DMaps help illustrate the effects (positive or negative) of quality concerns on each other and their direct, enabling, or systemic impacts over time. The Sustainability Quality Model (SQModel) helps define quality concerns and their sub-characteristics, per dimension. It also helps to define their measurement metrics, which can be used to define Key Performance Indicators (KPIs). These KPIs can help formalize the measurement process, making it actionable by linking high-level goals to concrete measurements and actionable tactics. As fairness is often represented as a collection of quality concerns (e.g., transparency, accessibility, bias, etc.), its cross-cutting nature can be represented by using an architecture perspective~\cite{rozanski2012} which is designed to facilitate representing cross-cutting quality concerns. 

\section{Design Thinking Approach}
\label{sec:approach}
\begin{figure*}[!htbp]
    \centering
    \includegraphics[width=0.9\linewidth]{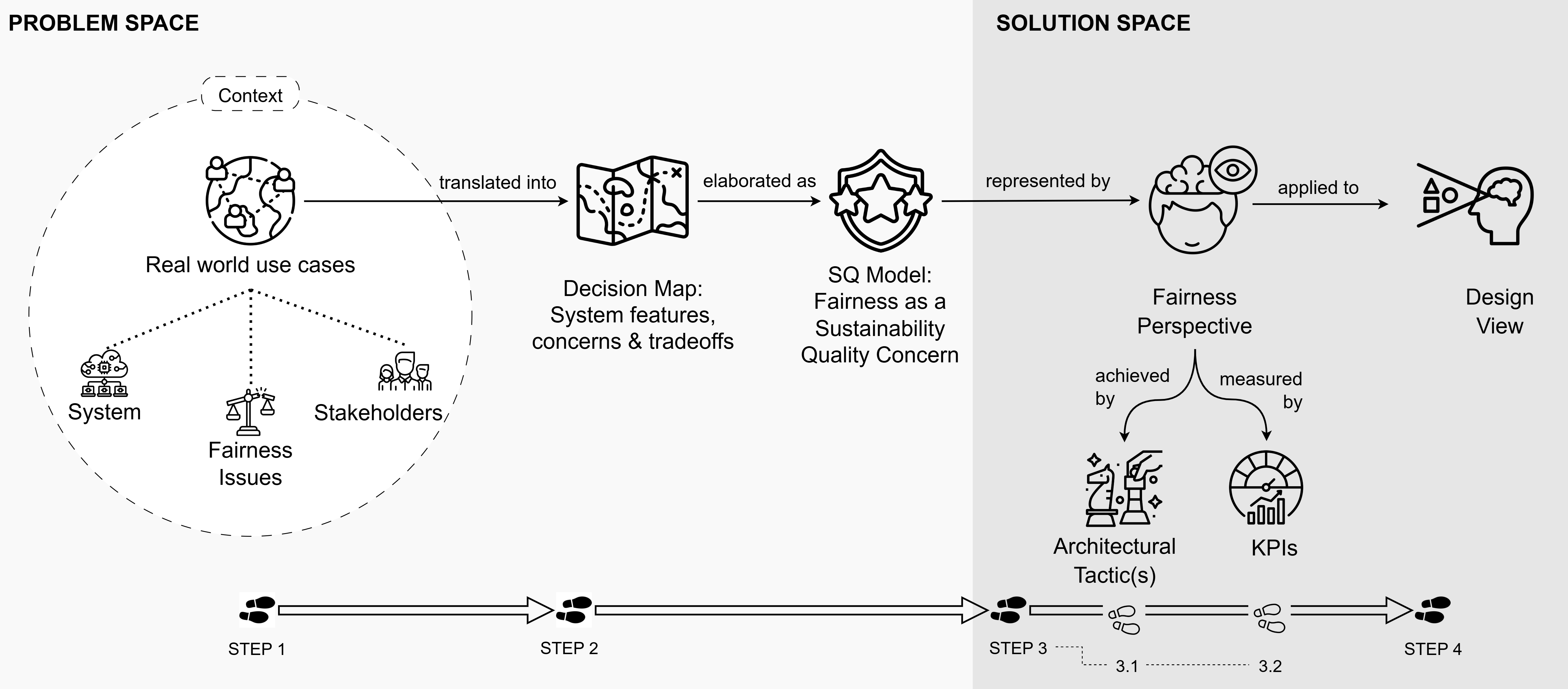}
    \caption{Overview of the DT Approach}
    \Description{Overview of the DT Approach}
    \label{fig:overview}
\end{figure*}

DT refers to a human-centered problem-solving approach that emphasizes understanding stakeholder concerns, their context, redefining problems, and creating innovative solutions that fulfill stakeholder needs.
Our DT approach aims to facilitate the identification of fairness issues and to find actionable and measurable solutions to integrate fairness in SA design. Hence, we divide our DT approach into a problem space and a solution space. Figure \ref{fig:overview} shows an overview of our DT approach. 

\textbf{Problem Space.}~
Exploring the problem space involves identifying the problem of fairness and reasoning about fairness in terms of associated quality concerns. The steps are as follows.

\textit{Step 1 -- } It involves the identification of a real-world use case on fairness issues in a digital system, its stakeholders, and the context.

\textit{Step 2 --} It involves the exploration of the problem space using a DMap and the SQModel from the SAF Toolkit~\cite{Lago2024-SAF-Toolkit}. The DMap helps translate the system into relevant features, quality concerns, their effects on each other, their impacts over time, and sustainability dimensions. These effects help identify trade-offs between different quality concerns. Next, the SQ Model elaborates on the identified quality concerns by defining them in the context of the use case and as per sustainability dimensions. The final SQ Model provides an overview of fairness as a sustainability quality concern, represented by a collection of quality attributes (QAs).  

\textbf{Solution Space.}~
Exploring the solution space involves representing fairness using a \textit{fairness perspective}, applied through a design view. The next steps are as follows.

\textit{Step 3 -- } It involves the creation of a fairness perspective. To capture a set of elements satisfying a particular fairness concern, the fairness perspective is adapted from the \textit{Sustainability Perspective} by Funke and Lago~\cite{FunkeLago_Perspectives_2025}, which is rooted in the general concept of architectural perspectives~\cite{rozanski2012}. Our \textit{fairness perspective} consists of the following elements: (i) \textit{Desired quality} which defines fairness as a quality of the system under consideration, (ii) \textit{Quality concerns} which are scoped to the fairness definition, (iii) \textit{Architectural tactics} which, if implemented, would lead to achievement of the desired quality, (iv) \textit{Fairness Indicator} which is a KPI that can be used to measure and monitor the level of fairness in the system, and (v) \textit{Problems and Pitfalls} which are the issues which may hinder the achievement of the fairness. 

\textit{Step 3.1 -- } Architectural tactics from the fairness perspective are documented using the template from the Awesome and Dark Tactics (AADT) archive\footnote{\url{https://s2group.cs.vu.nl/AwesomeAndDarkTactics}}. 

\textit{Step 3.2 -- }KPIs from the \textit{fairness perspective} are documented using the KPI template~\cite{KPI-Fatima-2024}. Quality concerns from the fairness perspective and metrics identified as part of the SQ model help to build a KPI that meets the fairness goal for the desired quality in the fairness perspective.

\textit{Step 4 -- } Finally, a fairness-based design solution is illustrated by using a design view, i.e., an architecture diagram. 

Our DT steps align with the five stages of DT~\cite{dSchool2009}. Steps 1 and 2 help elicit human-centered fairness concerns (empathize), Step 3 refines them into a sustainability perspective (define), Step 3.1 explores architectural tactics and options (ideate), Step 3.2 proposes measurable criteria to monitor (and test) the interventions represented by Step 4 as design views that operationalize fairness (prototype). Due to the limited duration of the course, the design is not tested via the implementation of the prototype. 

\section{Study Design and Execution}
\label{sec:design}
The study was carried out as part of a graduate-level Fair Digital Design that spanned a period of seven weeks, where the student attended a lecture and a working session per week. A total of 26 students took the course. The students worked in pairs on the \textit{Design for Fairness (D4F)} assignment. The DT approach was executed as part of this assignment. Each group explored a fairness issue in the domain of cloud/edge data management. The students were free to choose a fairness issue of their choice. As a final output, they were expected to present the solution as a fairness perspective and design views. Our study focused on the \textit{design} and \textit{reflections} of the DT approach, applied as part of the D4F assignment. During this course, we introduced our DT approach and operationalized learning through a structured assignment. One lecture was dedicated to teaching the DT approach and the relevant techniques to be used in the assignment. During the working sessions, a facilitator guided students on the techniques taught in the lecture and how to use them for the assignment. The students also received feedback on their progress during these sessions. The graded assignments with corrections were used as a source for data extraction and analysis.

We applied the following inclusion criteria to ensure the quality of the data under analysis:
\newline
(i) Data used for this study is extracted from assignments with a grade $\geq 60\%$.
\newline
(ii) Data used for this study is referenced using the peer-reviewed literature.    
We selected all 13 assignments as they met the inclusion criteria. Data extraction was performed for each step of the DT approach. 

We provide an online replication package~\cite{rep-pkg} containing templates for each DT step and the extracted data, including only assignments with students' consent (i.e., all except A12).

\textit{\textbf{Step I - Use case:}}~
We used \textit{open coding}~\cite{strauss1990open} for data extraction. This resulted in 11 elements of the use case description (see Table \ref{tab:usecase-cats}). The data from the use cases was extracted into these categories.
Further, we categorized fairness issues based on the type of fairness issue described in the use case problem descriptions. This categorization was based on the fairness definitions from philosophy theory, described in the following. 

\greybox{
\textbf{Distributive Fairness -- }Resources, benefits, and burdens should be allocated among individuals or groups based on fair and just criteria such as equity, equality, or need. These allocations should be made to ensure fairness in outcomes, focusing on transparent, consistent, and just distribution principles that stakeholders recognize as legitimate~\cite{colquitt2001}.
}

\greybox{
\textbf{Procedural Fairness -- }Decision-making processes should be built and applied consistently, transparently, and without bias. It must be ensured that all affected parties have a voice, that rules are followed, and explanations are provided to ensure fairness in the process~\cite{leventhal1980procedural}.
}

\greybox{
\textbf{Interactional Fairness -- }Individuals should be treated with dignity, respect, and honesty during interactions, with clear, truthful, and timely communication that reflects consideration of their concerns~\cite{bies1986interactional}.
}

These definitions were matched with the descriptions of use cases and the fairness problems. The categorization was carried out independently by two co-authors and later compared to resolve differences and finalized after mutual consensus. 
\begin{table}[!htb]
    \centering
    \small
    \caption{Use Case Template}
    \begin{tabular}{p{2.5cm}p{5.2cm}}
    \toprule
        \textbf{Element} & \textbf{Description} \\ \midrule
        Domain & Describes the general area where the system operates (e.g., healthcare, education, etc.)\\ \hline
        Actor(s) & Stakeholder(s) of the system who interact with it or are affected by it \\ \hline
        System name & An identifiable name of the system \\ \hline
        System description & A brief overview of the system's domain, prominent features in the context of the use case, and its intended purpose \\ \hline
        Use case focus & Specific tasks performed by the system/actors\\ \hline
        Problem in the context of the actor & Challenges that the actors face due to the system operating in a certain way \\ \hline
        Problem in the context of the system & Challenges that the system faces to operate in a certain way \\ \hline
        Technical subject of fairness issue & The technical component where fairness is at risk \\ \hline
        Impact of the design issue on the actor & The impact of the design choices on the actors in terms of fairness \\ \hline
        Impact of the design issue on the system & The impact of the design choices on the system in terms of fairness \\ \hline
        Proposed solution & The solution to address fairness issues in the system at the architectural level \\ \bottomrule
    \end{tabular}
    \label{tab:usecase-cats}
\end{table}

\textit{\textbf{Step 2 - Decision Map and Sustainability Quality Model:}}~
We employed \textit{a priori coding}~\cite{bingham2022apriori} for data extraction, as this step required the use of templates from the SAF Toolkit, which has predefined elements that we used as categories for data extraction. Once the data was coded, we performed further categorization using open coding to gather insights about possible patterns.
Each SQ model contained a subset of quality concerns from the Decision Map, defined according to the mapped sustainability dimension. To identify QA trade-offs and synergies, we used this common subset of QAs and observed the effects in the DMaps as \textit{positive, negative, or neutral} over time as\textit{ immediate, enabling, or systemic} impacts. 
The SQ Models helped define QAs in a certain sustainability dimension. 
We extracted data from the SQ Models and grouped it by dimensions. In each dimension, we categorized the QAs into different QA categories. For the technical dimension, we use the ISO/IEC 25010 standard\footnote{\url{https://www.iso25000.com/index.php/en/iso-25000-standards/iso-25010}}. For other dimensions, we performed deductive coding to identify categories, as no standard is currently in place.

\textit{\textbf{Step 3 -- Fairness Perspective:}}~
We employed \textit{a priori coding}~\cite{bingham2022apriori} for data extraction. This choice was made because this step required the use of templates from different frameworks(such as the Sustainability Perspective, AADT Template, and KPI Template), all of which have predefined data categories for data extraction.
The fairness perspective defines fairness as a desired quality. We analyzed these definitions in combination with tactics and KPIs to observe similarities and differences, specifically for similar domains. We also compared the implications of the definitions with the types of fairness assigned to the use cases. This was to observe any changes in perception of fairness as we transition from the problem to the solution space. We extracted all tactics from the fairness perspectives and mapped them across fairness types based on tactic intent. This served as a guide to indicate the type of fairness the tactic supports, i.e., distributive, procedural, or interactional. 
The KPIs from the fairness perspective were elaborated using the KPI framework, which outlines the goal, the critical success factor, the KPI with its goals and actions (if the target is not achieved), the metrics, and their corresponding measures. We analyzed the data in these categories to see similarities, patterns, and insights to enable fairness in different contexts and for different fairness goals.

\textit{\textbf{Step 4 -- Design View:}}~
We used \textit{directed content analysis}~\cite{hsieh2005} by starting with the codes from the work of Migliorini~\etal~\cite{Migliorini2024} followed by developing additional categories inductively from the data to capture concepts not addressed in the previous framework.  
We also looked at the elements of the views and analyzed them in light of the problem context, the provided solution, and the fairness types. Using these sources of information, we determined which elements support a specific type of fairness need. We also identified new elements from the view data and provided some recommendations in the reflections. 

The choice of inductive coding for steps 1 and 4 (part) was because the students were given creative liberty to elicit these steps, hence, information was documented in different formats. The choice of inductive coding was motivated by the aim of finding practical and novel ways to represent fairness issues and their solutions. 
\section{Results}
\label{sec:results}

Our results show that the students focused on use cases specifically in the domains of healthcare (A02, A06, A10, and A12), education (A08 and A11), disaster management (A01), agriculture (A04), entertainment (A05), finance (A09), and hospitality (A13). Two use cases are domain-independent and target cloud software in general (A03 and A07). Table~\ref{tab:re-usecase} shows an example use case represented using the use case template created during the data extraction process. The details of the other use cases are available in our replication package~\cite{rep-pkg}. 
\begin{table}[!htbp]
    \centering
    \small
    \caption{Example: Step 1 -- Usecase (A08)}
    \begin{tabular}{@{\hspace{0.05cm}}>{\raggedright}p{2.4cm} p{5.4cm}}
    \hline
    \textbf{Element} & \textbf{Description} \\ \hline
        Domain & Education \\ \hline
        Actor(s) & High school student in a remote location in a country in the Global South \\ \hline
        System name & Online education platform \\ \hline
        System description & The educational platform is hosted on a global cloud infrastructure with an edge network, offering online classes for high-school students. It offers digital resources, on-demand and real-time online classes, and an assessment component, offering self-paced and real-time proctored tests. \\ \hline
        Use case focus & Uninterrupted real-time online classes \newline Uninterrupted virtual tests  \\ \hline
        Problem(s) in the context of the actor & Network fluctuations cause delays, video lag, and disconnections, disrupting student exams, not because of their ability, but because of poor connectivity. Network issues lead to session interruptions, significantly impacting students' learning experience and progress. \\ \hline
        Problem(s) in context of the system & Poor connectivity impacts real-time data transfer in the education platform \\ \hline
        Technical subject of fairness issue & Data management over unreliable networks \\ \hline
        Impact of the design issue on the Actor & Student's learning experience and progress. \\ \hline
        Impact of the design issue on system & Not provided \\ \hline
        Proposed solution & The architecture will address network issues and aim at counter-balancing the disadvantage caused by socioeconomic circumstances, with features designed to ensure fair access. \\ \hline
    \end{tabular}
    \label{tab:re-usecase}%
    \end{table}

In the following, we present each finding from our results with a concise statement summarizing the primary data pattern, followed by a detailed explanation, and a concluding takeaway. 

\begin{figure}[!htbp]
    \centering
\includegraphics[width=0.6\linewidth]{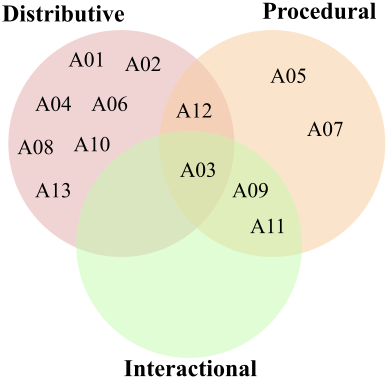}
    \caption{Overlapping use cases per fairness type}
    \Description{Figure showing overlapping use cases per fairness types}
    \label{fig:venn-fairness-types}%
\end{figure}

\noindent\textbf{Distributive fairness issues dominate in problem descriptions; procedural and interactional issues are less frequent.}
\newline
The classification of fairness issues in terms of fairness types shows a mix of disjoint and overlapping cases (see Figure~\ref{fig:venn-fairness-types}). 
In particular, we observe that most issues are described as distributive fairness issues (in A01, A02, A03, A04, A06, A08, A09, A10, A12, A13), six as procedural (in A05, A07, A09, A11, A12), and three ( in A03, A09, A11) as interactional (see Table~\ref{tab:fairness-classification} for issue descriptions). 

Only one issue (in A12, \textit{a healthcare diagnostic system}) is classified as both distributive and procedural because the design of the diagnostic algorithm and the choice of training data are procedural issues in the context of the system, while the impact of bias leading to unequal treatment is a distributive fairness issue in the context of the user. Similarly, two issues (in A09 and A11) are classified as both procedural and interactional. In A09 (credit scoring system), there is a lack of transparency in the process used for credit scoring. This, coupled with the lack of explainability, makes it both a procedural and an interactional issue. In A11 (education record verification system), there is no procedure in place to access and get information about unavailable diplomas, making it a procedural issue, whereas the lack of information to the user is a communication issue, making it interactional in nature. 
\newline
Only one issue (in A03, \textit{an AI job scheduler}) is classified across all fairness types. The inability to allocate the burden of scheduling jobs with energy-efficient resources is a distributive fairness issue. The lack of a sustainability-aware scheduling mechanism is a procedural issue, whereas for the developers, this is an interactional fairness issue, as it hinders them from making informed decisions based on transparency.

In summary, in our data, the issues of distributive fairness focus on unequal access to resources based on geographical differences or varying data storage policy decisions. The issues of procedural fairness focus on the lack of control or transparency about the process through which the software operates and handles data, sometimes leading to disparities and unfair outcomes. The issues of interactional fairness focus on the lack of explainability of the decision-making process, the lack of transparent/poor communication with users, harming user trust, and the usability experience. 

\greybox{
\textit{\textbf{Takeaway -- }}Distributive fairness issues appeared more frequently, possibly because they are based on outcomes and manifest outside the system, making them more noticeable to the students. To support the identification of different types of issues in future editions of the course, we aim to provide an overview of fairness types and their possible manifestations in digital systems.
}
\begin{table}[!htbp]
    \centering
    \small
    \caption{Fairness Issues mapped on Fairness Types}
    \begin{tabular}{|@{\hspace{0.05cm}}m{0.35cm}|
@{\hspace{0.08cm}}p{6.6cm}@{\hspace{0.08cm}}|
    @{\hspace{0.05cm}}m{0.13cm}|
    @{\hspace{0.05cm}}m{0.13cm}|
    @{\hspace{0.05cm}}m{0.13cm}|}
    \hline
        
        \textbf{ID} & \textbf{Issue Description} & \textbf{D} & \textbf{P} & \textbf{I} \\ \hline

        A01 & Marginalized communities have limited access to disaster response tools, leading to unequal protection. & \ding{108} & ~ & ~ \\ \hline
        A02 & AI predictions seem less accurate for patients from minority groups due to imbalanced data. & \ding{108} & ~ & ~ \\ \hline
        A03 & ML engineers have no control/information over sustainability-aware job scheduling, indicating a lack of voice and transparency. & \ding{108} & \ding{108} & \ding{108} \\ \hline
        A04 & Lack of real-time availability of decision-making information in rural areas & \ding{108} & ~ & ~ \\ \hline
        A05 & Users cannot transfer data between music providers & ~ & \ding{108} & ~ \\ \hline
        A06 & Patients from rural areas have trouble connecting to the system, hindering remote patient health monitoring. & \ding{108} & ~ & ~ \\ \hline
        A07 & Privacy and legal protections under GDPR are difficult to apply equally across regions. & ~ & \ding{108} & ~ \\ \hline
        A08 & AI-driven education systems are less accessible to under-resourced learners, widening the learning gap. & \ding{108} & ~ & ~ \\ \hline
        A09 & Credit reductions are made by opaque algorithms without user input or an appeal channel. & ~ & \ding{108} & \ding{108} \\ \hline
        A10 & Inefficient cloud resource allocation causes long patient waiting times in healthcare. & \ding{108} & ~ & ~ \\ \hline
        A11 & Diploma access processes are unclear, reducing perceptions of fairness. & ~ & \ding{108} & \ding{108} \\ \hline
        A12 & Biased diagnostic algorithms result in unequal patient treatment. & \ding{108} & \ding{108} & ~ \\ \hline
        A13 & AI support distribution is uneven between urban and rural areas. & \ding{108} & ~ & ~ \\ \hline
        \multicolumn{5}{l}{\footnotesize{Fairness Types; D: Distributive, P: Procedural, I: Interactional}}%
    \end{tabular}%
    \label{tab:fairness-classification}%
\end{table}

\begin{figure}[!htbp]
    \centering
\includegraphics[width=\columnwidth]{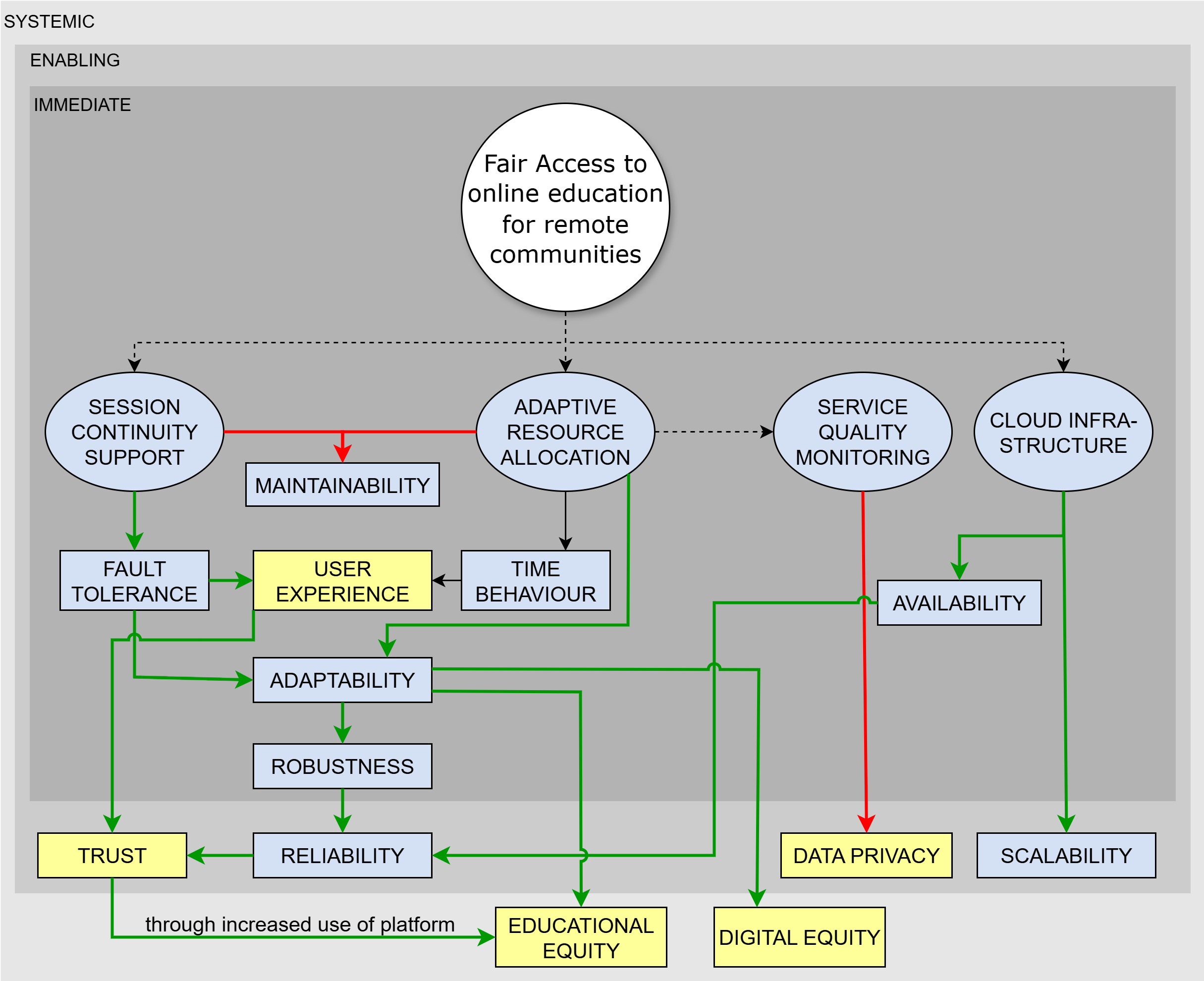}
    \caption{Example: Step 2 -- DMap (A08)}
    \Description{Figure showing a DMap as an example from use case A08}
    \label{fig:re-dmap}
\end{figure}

\noindent\textbf{Technical quality concerns can cause immediate impacts that ripple into systemic social and economic effects over time.}
\newline
DMaps show information on QA trade-offs and synergies that are rooted as direct impacts, but over time lead to enabling and systemic impacts. For example, in the use case A08 (\textit{Domain: Education}), the implementation of the adaptive resource allocation feature (see associated Figure~\ref{fig:re-dmap}) negatively affects maintainability, as it increases technical complexity and effort. However, when observed over time, this feature contributes positively to adaptability, creating a positive effect on robustness (T)\footnote{(T): Technical, (En): Environmental, (Ec): Economic, (S): Social} as a direct effect. This robustness further enables reliability (T) and fosters trust (S), over time. Establishing trust (S), in turn, can have a systemic effect in promoting educational equity (S).

In the use case A07\footnote{See Replication Package~\cite{rep-pkg} for DMap} (\textit{Domain: Cloud Software -- AWS}), data sov\-er\-eign\-ty (T) is positively reinforced as an immediate impact of GDPR compliance. This, in turn, has a positive enabling effect on regulatory compliance (S), which leads to a positive systemic impact on third-party compliance (Ec). This illustrates how a technical intervention can produce long-term effects in the social and economic dimensions of sustainability, contributing to fairness over time through systemic influence. Since the positive systemic impact lies within the economic dimension, it creates a strategic incentive for businesses to adopt fair data management policies.
\greybox{ 
\textit{\textbf{Takeaway -- }}Technical interventions can drive long-term social and economic fairness, providing strategic incentives for businesses to adopt fair practices. A fairness-first DT approach may enable integrating interventions for social quality concerns early at design time. 
}

The classification of data from the SQ Model shows that out of 148 QAs, 46.6\%  (69) QAs were classified as technical, 6.8\% (10) as environmental, 12.8\% (19) as economic, and 33.8\% (50) as social. 
The SQ Model also contains metrics that can be used to quantify some QAs. Although our study is specific to fairness, some of the provided metrics are agnostic of the general goal of measurement and, therefore, can be reused for quality monitoring in other contexts. The information from the SQ Models is later used in defining the fairness perspective. 
Both QA definitions, their mapping on dimensions, and metrics can be found in the replication package~\cite{rep-pkg}.

\noindent\textbf{Fairness definitions vary within similar domains and in granularity levels.}
\newline
We observed certain conflicts in how fairness is defined in different fairness perspectives. For example, A07 emphasizes strong privacy protection in order to adhere to the principles of data sovereignty and compliance with GDPR. In contrast, A08 stresses the sharing of resources for equitable access to educational resources, which may cross continental boundaries and cause data sovereignty and privacy issues. 
We also see differences in the definition of desired quality within a domain. An analysis of healthcare use cases (A02, A06, A10, A12) shows that fairness concerns in this context are multifaceted and often intersect across the categories of distributive and procedural fairness. A02, initially categorized as distributive, emphasizes fairness in data-driven decision making, particularly through transparency, explainability, and equitable treatment across demographics. This reinforces its alignment with interactional fairness while also addressing distributive concerns through accessibility. A06, with its comprehensive focus on representation management, data quality, connectivity, and compliance across locations, bridges distributive fairness (equal access and representation) and procedural fairness (mechanisms ensuring fairness in data handling and system operations). A10, concerned with equitable access to data resources, performance metrics, and adherence to international standards, reflects both distributive fairness (resource access) and procedural fairness (standardized measurement and compliance frameworks). Collectively, these cases underscore that fairness must be understood both as a system outcome and as the mechanisms by which such outcomes are produced.

Our results show that fairness definitions vary in granularity from a very high level (fair and sustainable data management in A01) to very specific (secure music data porting in A05). Although quality concerns are identified as economic in ten SQ Models, we still see a scarcity of a fairness perspective that includes economic aspects, except for A01 and A10. These two fairness perspectives include regulatory compliance and cost efficiency. 

\greybox{ 
\textit{\textbf{Takeaway -- }}The definition of fairness is driven by context and stakeholder needs. What counts as fair in one system under certain constraints may not be considered fair for another. These trade-offs in digital systems should be understood both in terms of system outcomes and processes that produce them. 
}

\noindent\textbf{Architectural tactics serve as procedural fairness-based interventions.}
\newline
Our results show that all tactics fall under procedural fairness because the intrinsic nature of tactics being procedural interventions. 
Four tactics fall under both procedural and interactional categories, depending on whether the solution is system-facing or user-facing. For example, regulatory environments (e.g., EU AI Act, GDPR) increasingly demand justification and transparency. For architects to know how explainability serves different fairness roles would help in meeting such requirements strategically through procedural fairness-based interventions. Although some fairness problems were framed as distributive (such as A01 and A10), the fairness intent in the architectural tactics used to solve them now includes some procedural elements. Interactional fairness was the least addressed by architectural tactics. Table~\ref{tab:tactics} shows all tactics categorized according to the three types of fairness.

\greybox{ 
\textit{\textbf{Takeaway -- }}
Overall, we positively noted that although not all fairness types were part of the problem space, as the solution space evolved, we saw a diversity of fairness types being addressed through the fairness-first DT approach by the use of tactics.
}

\begin{table}[!htbp]
    \centering
    \small
    \caption{Architectural Tactics Mapped to Fairness Types}
    \begin{tabular}{|@{\hspace{0.08cm}}p{7.1cm}|
    @{\hspace{0.05cm}}m{0.12cm}|
    @{\hspace{0.05cm}}m{0.12cm}|
    @{\hspace{0.05cm}}m{0.12cm}|}
    \hline
        \textbf{Architectural Tactic (Use case ID)} & \textbf{D} & \textbf{P} & \textbf{I} \\
        \hline
        Smart Data Tiering (A01) & & \ding{108} & \\ \hline
        Adaptive Resource Allocation (A01, A08, A13) & \ding{108} & \ding{108} & \\ \hline
        Green Computing Strategies (A01) & & \ding{108} & \\ \hline
        Explainable AI (A01, A02, A09) & & \ding{108} & \ding{108} \\ \hline
        Regulated Storage Access (A01) & \ding{108} & \ding{108} & \\ \hline
        Realtime Bias Detection (A02, A12) & \ding{108} & \ding{108} & \\ \hline
        Adaptive Learning Mechanisms (A02) & \ding{108} & \ding{108} & \\ \hline
        Carbon footprint aware scheduling (A03) & & \ding{108} & \\ \hline
        Data center selection using adaptive desirability functions~(A03) & \ding{108} & \ding{108} & \\ \hline
        Runtime process migration to suitable data centers (A03) & \ding{108} &  \ding{108}& \\ \hline
        Monitor environment of individual jobs (A03) & & \ding{108} & \\ \hline
        Local Data Processing via Edge Computing (A04) & & \ding{108} & \\ \hline
        IoT Sensor Network for Real-Time Data Collection (A04) & & \ding{108} & \\ \hline
        Distributed Edge Devices (A04) & & \ding{108} & \\ \hline
        Blockchain for Transparency (A04) & & \ding{108} & \ding{108} \\ \hline
        Standardized data portability between platforms (A05) & & \ding{108} & \ding{108} \\ \hline
        Adaptive Edge Processing Prioritization (A06) & \ding{108} & \ding{108} & \\ \hline
        Data Access Control for GDPR (A07) & \ding{108} & \ding{108} & \\ \hline
        Prioritized Traffic Management (A08) & & \ding{108} & \ding{108} \\ \hline
        Edge Caching (A08) & \ding{108} & \ding{108} & \\ \hline
        Provenance-Driven Transparency (A09) & & \ding{108} & \\ \hline
        Tiered Storage (A10) & \ding{108} & \ding{108} & \\ \hline
        Energy-aware Scheduling (A10) & & \ding{108} & \\ \hline
        Compress infrequently accessed data (A10) & & \ding{108} & \\ \hline
        Continuous monitoring \& mitigation of AI bias (A12) & \ding{108} & \ding{108} & \\ \hline
        Cloud-based retraining with diverse datasets (A12) & \ding{108} & \ding{108} & \\ \hline
        \multicolumn{4}{l}{\footnotesize{Fairness Types; D: Distributive, P: Procedural, I: Interactional}}%
\end{tabular}
\label{tab:tactics}%
\end{table}
%

\noindent\textbf{KPIs help turn fairness goals into measurable outcomes.}
\newline
Architectural tactics contribute to making fairness actionable in SA design, however, it is important to monitor the impact of these interventions over time. Our DT framework assists in creating KPIs for fairness monitoring. 

Analysis of KPI data shows that some measurements are performed by collecting data from the digital system, hence classified as \textit{internal}. For others, measurements are performed by collecting data from users, therefore, categorized \textit{external}. Some KPIs use a mix of internal and external measures and are therefore classified as \textit{hybrid}.
Six KPIs were classified as internal (A03, A12, A01, A06, A10, and A13), while others were classified as hybrid (A02, A07, A11, A05, A09, A04 and A08). 
Among the KPIs, we observe two distinct approaches to setting targets (i) \textit{single value targets} and (ii) \textit{conjunctive targets}. In the first approach, various underlying metrics are aggregated into a single numerical value using a mathematical function. This simplifies monitoring and decision-making at the macro level, offering a unified view of performance. However, such aggregation can obscure the contribution of individual metrics, making it difficult to understand why a particular KPI value was achieved.
In contrast, the conjunctive approach treats each metric separately with its own threshold. Instead of single-valued functions, these KPIs use logical functions. A KPI is considered successful only if all KPI metrics meet their respective thresholds. This approach improves transparency and provides detailed information, allowing targeted interventions through KPI actions. However, this approach can compromise the benefits of simplification and abstraction that single-value KPIs provide, potentially making high-level comparisons and reporting more complex.

\greybox{ 
\textit{\textbf{Takeaway -- }} The choice between single-value vs. conjunctive KPIs involves a trade-off between clarity for strategic decision-making and granularity for operational insight. The ultimate choice lies in the monitoring goal. 
}

We observed that some KPI targets were based on architectural tactics from a fairness perspective. This shows that architectural interventions can facilitate the achievement of fairness goals, can be monitored over time, and are well-handled early in the software development lifecycle. 
For example, in A01, actions to improve the fairness index included tactics such as implementing smart caching mechanisms to optimize data retrieval speed and resource allocation adjustments to dynamically balance workloads. Other tactics as actions include adaptive resource allocation and processing on the edge, and load balancing. 
The KPI for A08 measures the continuity of the session, with the target of maintaining continuity above 95\%. This requires defining a performance requirement for sustained access under varying loads. Meeting this requires the use of adaptive resource allocation, which guides architectural tactics such as auto-scaling and real-time monitoring. This chain from KPI to the design tactic illustrates how high-level goals can be operationalized within the system architecture and monitored over time.

\greybox{ 
\textit{\textbf{Takeaway -- }}
KPIs act as a bridge between fairness ideals and system design, translating fairness into measurable metrics without losing its ethical purpose.
}

\noindent\textbf{Contextual design view elements help embed fairness in SA.}
\newline
Our results show that some design views follow notations from standards like ArchiMate\footnote{\url{https://www.opengroup.org/archimate-forum/archimate-overview}} and UML\footnote{\url{https://www.omg.org/uml/}}, while others use a mix of notations from different architecture viewpoints (such as deployment, functional, operational, etc.) and some custom notations specific to the context of the use case. 

The level of granularity is high for most views (A01, A02, A03, A04, A05, A06, A07), while medium for others (A09, A10, A11, A12, A13). Only one view had a low granularity level and was depicted as an activity diagram using an operational viewpoint for job scheduling. 
The views focus on representing the interaction of stakeholders with the features and the representation of the fairness issue through two types of interactions, (i) human--technology and (ii) technology--technology. 
The former included human actors (A05, A09, A13), user (A01), and stakeholders (A07, A08, A12). The latter included components of different systems communicating through interfaces (A01-A13). 

\begin{figure}[!htb]
    \centering
    \includegraphics[width=1\linewidth]{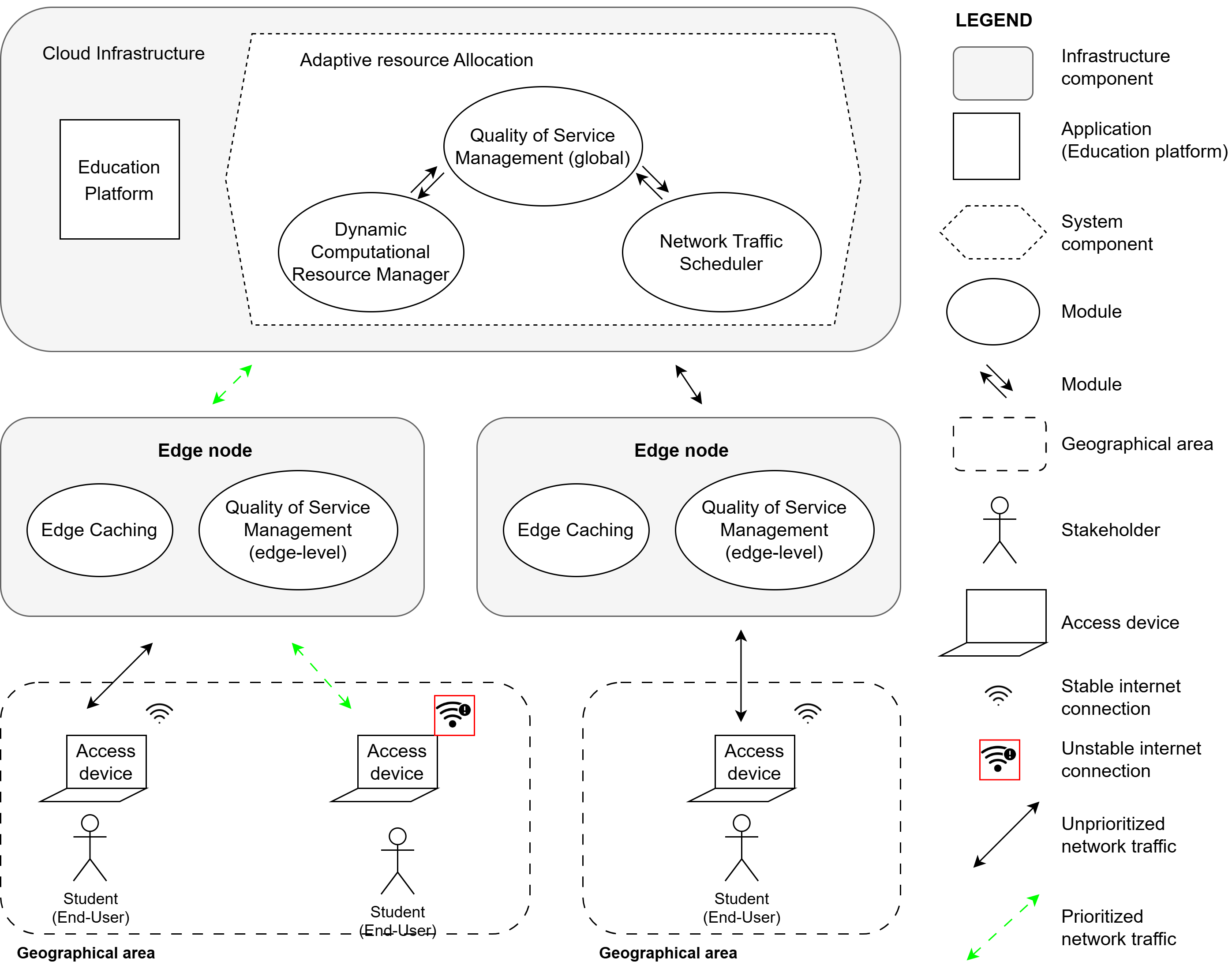}
    \caption{Step 4 -- Design View}
    \Description{Figure showing design view from use case A08 as an example}
    \label{fig:re-fv}%
\end{figure}

One design view was textual (A01), while the other 12 were graphical. The views provide an overview of the static and dynamic behavior of the architecture. The views provide an overview of the features and components, their interactions with each other, external systems, and stakeholders.
We also observe some elements in the design views that are not standard but rather improvised to represent the context (see Figure~\ref{fig:re-fv}). For example, the elements of the \textit{geographical area} (A08, A06) capture distributive fairness by reflecting the environmental and regional factors that influence the equitable allocation of resources. The elements of \textit{connection quality} and \textit{network traffic prioritization} (A08) further support distributive fairness by modeling the varying capabilities of the infrastructure and ensuring the priority of the data flow where necessary. The \textit{constraint} elements for privacy and other regulations (A02, A07) emphasize procedural fairness by embedding compliance mechanisms that ensure that decisions respect established rules and stakeholder rights. The \textit{stakeholder influence} relationship (A07) highlights interactional fairness, representing the impact of organizational actors and their power dynamics within the system. Finally, \textit{representation management} and \textit{adaptive monitoring} (A06) integrate procedural fairness through continuous system assessment and dynamic resource adjustment based on KPI thresholds.

\greybox{ 
\textit{\textbf{Takeaway -- }}Contextual elements help document the context and rationale behind design decisions within the design views, hence providing a mechanism for architecture knowledge preservation.
}

\noindent\textbf{Explicit representation of targets and actions in design views makes fairness-driven architecture adaptations traceable.}
\newline
In A02, the activity diagram integrates KPIs directly into the decision-making and scheduling logic, and metrics guide the dynamic selection/reassignment of data centers. This integration makes fairness logic, including the prioritization of user preferences and equitable job distribution, explicit within the operational flow and provides a traceable link between QAs and adaptive system behavior.

A06 presents a similar contextual connection for connectivity status, proposing a representation monitor to enable adaptive resource allocation. The design view incorporates the KPI for representation management and its thresholds, making explicit when actions should be triggered to maintain fairness once a threshold is breached. This explicit integration of monitoring, thresholds, and adaptive response within the view provides clear traceability from the fairness requirement to the architectural element that fulfills it.
A08 (see Figure~\ref{fig:re-fv}) uses boxes (connection stability, geographical location) and lines (prioritized data transmission) to represent contextual information that is key to ensuring fairness. It combines the underlying problem and the proposed solution in a single, self-contained view. The view shows how different edge nodes transmit data with network prioritization for rural regions with poor connectivity. The adaptive system component includes a dynamic computational resource manager, a network traffic scheduler, and adaptive resource allocation to maintain Quality of Service. The visibility of external interactions in specific situations, such as unstable connectivity, strengthens the soundness of the architecture design representation.

\greybox{ 
\textit{\textbf{Takeaway -- }}Representing  KPI targets and adaptive actions in design views helps fairness requirements become operational and traceable in SA. This helps strengthen the link between fairness goals and concrete interventions needed to achieve them.
}

\section{Reflections}
\label{sec:reflection}

We present our reflections on the insights about fairness in the context of SA and our fairness-first DT approach for education. 

\subsection{On Fairness in Software Architecture}
Our analysis revealed a consistent pattern where issues framed primarily as distributive fairness problems (in the problem space) were often addressed with procedural fairness–based solutions (in the solution space). For example, several use cases showed that concerns about distributive fairness (e.g., resource allocation) were resolved through procedural interventions across different use cases, including adaptive learning mechanisms (in A02), process transparency (in A04), and through monitoring components (in A08).

The DMaps helped us observe possible trade-offs between QAs as we explored the problem space. In addition, they helped to visualize the possible enabling and systemic impacts of technical features over time. Such exploration of the problem space, specifically for multidimensional concerns such as fairness, is essential as it allows for proactive management of overtime cross-dimensional impacts within the design phase instead of addressing them later.

We observed that explainability as a quality concern served a dual role. When designed for system-facing use, it enabled auditing and compliance (procedural fairness), while user-facing implementations supported trust and comprehension (interactional fairness). Hence, it is important to define a QA with its different interpretations based on the intent as per the dimension(s) it supports, which is facilitated by the SQ Model in our DT approach. 

By applying fairness perspectives on design views, we established a clear chain to (i) trace architectural elements to the QA(s) they affect, (ii) the KPIs that measure fairness using these QA metrics, and (iii) the tactics that make these KPIs actionable and help enable the desired level of fairness. Building on this \textit{traceability chain}, we suggest extending sustainability perspectives (i.e., architectural perspectives) with the element of KPIs. Incorporating KPIs within a perspective enables systematic monitoring of the desired qualities. The KPI definition in the problem space becomes especially helpful for new and unexplored concerns like fairness or sustainability, in general. KPIs at this stage facilitate the collection of relevant data, which in turn enables informed design decision-making and the reconsideration of earlier choices. Over time, as sufficient data is collected, such KPIs can evolve beyond the architecture problem space and inform, e.g., architecture guidelines or patterns with clear targets for future solutions.

In the solution space, many design view elements span multiple types of fairness, reflecting the cross-cutting nature of fairness concerns in SA. For example, stakeholders appear in all three types but play different roles, such as resource recipients (distributive), participants in decision processes (procedural), and participants in social interactions (interactional). It is important to consider these fairness types in hindsight when creating design views, while being careful with the intent of each choice. 

Complex concerns such as fairness require a composite view, as they do not fit neatly into one architecture viewpoint due to their multidimensional nature. Addressing a quality concern for fairness may require the representation of multiple elements (such as data, behavioral, stakeholders, deployment, and others) in a single view. This requires combining viewpoints or their elements, which can be achieved by using composite viewpoints. Composite viewpoints integrate components from various viewpoints, allowing architects to concentrate on concern-specific elements that extend across multiple viewpoints~\cite{rozanski2012}. Hofmeister \etal~\cite{hofmeister2007} emphasize that viewpoints should be flexible enough to facilitate composition or extension to address new or evolving concerns, which in our case is \textit{fairness}. The ISO/IEC/IEEE 42010:2011 standard~\cite{iso42010} also allows for creation of \textit{`custom viewpoints'} to address concerns that are not adequately supported by existing ones. Composite viewpoints can enable one to tackle multiple fairness types in concise views. Considering the multidimensional and cross-cutting nature of fairness as a quality concern, we recommend the use of composite views. 

Sections \ref{sec:bg} and \ref{sec:rw} show that context is key to understanding and solving fairness issues, and it is also a key element in the DT process.
Therefore, design views for fairness concerns must explicitly incorporate contextual elements. The design views in this study used elements such as geographic location, connectivity status, regulatory constraints, and stakeholder influences. Such contextual elements are essential to capture the environmental and social conditions that impact fairness-based decisions in the system. Including such contextual information allows architects to visualize the influence of external factors and stakeholder relationships on fairness outcomes. Furthermore, it preserves the rationale for making a decision within the design view. 

The presented DT approach provides a thinking framework to expand on fairness issues; facilitates understanding them in the context of system features and related concerns; and helps define a fairness perspective to apply fairness-supporting interventions during the design of SA. This helps ensure that fairness is treated as a priority concern and not retrofitted later into the system. 

\subsection{On Fairness-First Design Thinking in Education}

To understand and analyze the problem context, students need an understanding of the theory of fairness. Our results show that even when the initial problem focused on a single type of fairness, using the DT approach, aspects relevant to other fairness types appeared while exploring the solution space. Planning these aspects early can improve the effectiveness of solutions.  

Our experience showed that the students found it difficult to concisely define the use cases. Our results provide a use case template that facilitates the description of the problem in the context of the stakeholder and in the context of the system. A system designed to produce fair outcomes may still violate some fairness principles that affect the stakeholder(s). In contrast, economic, technical, or legal restrictions can hinder the implementation of a fairness-based design. Addressing both aspects of the issue (if applicable) is important to fully understand the fairness issue and its context. We aim to encourage students to use composite views and viewpoints for the representation of a design solution. We plan to integrate the learning from this research into the course and assignment design.

\section{Related Work}
\label{sec:rw}
This section discusses related work and outlines the research gaps bridged by our DT approach. 

\subsubsection*{\textbf{Design Thinking in Software Engineering Education.}}

The literature highlights the use of DT in teaching mobile application development~\cite{Valentim2017, Gama2018}. DT has also been used for project-based learning, requirement elicitation, software design, and development in a software engineering course~\cite{Palacin-Silva2017} and to teach the design of safety-critical software with physical hardware components~\cite{Cleland-Huang2017}. 

These studies collectively demonstrate that DT is a flexible framework, adapted to diverse educational contexts and problem spaces. However, the literature does not explicitly address the application of DT to SA, which is a fundamental phase for the construction of any software~\cite{bass2012software}.

\subsubsection*{\textbf{Fairness Considerations in Software Architecture}}

Shelly~\etal \cite{Shelley2012Fairness} question the assumption that technology is inherently neutral, suggesting that design choices can introduce unfairness into systems. This highlights the relevance of identifying fairness concerns during the early stages of design before they become embedded in the software solution. 

Fairness presents socio-technical challenges, requiring a perspective with professionals from both social and technical domains. To this aim, Ryan~\etal~\cite{Ryan2023} carried out a study with experts in machine learning and human-computer interaction, revealing that fairness is perceived in various and sometimes contradictory ways. Their work highlights data quality and context definition at design time as key challenges in achieving fairness, emphasizing that the fairness evaluation measurement process must be user-centric. In our study, we present a use case template to capture this context, together with an SQ Model that captures contextual definitions of quality concerns and the metrics to help measure them. Although context is important, the abstraction of the social context while designing sociotechnical systems can also lead to fairness issues. For example, Selbst \etal~\cite{Selbst2019} present four context-centered pitfalls in the design of such systems, some of which we avoid through our DT framework. These include the use of decision maps that help identify \textit{ripple effects} and leverage the SQ model to avoid \textit{the formalism trap} by defining quality concerns based on context per dimension (social, economic, environmental, and technical).

Elements of software design can also be leveraged as a tool to define fairness concerns early in the life cycle of software development ~\cite{santos2025softwarefairnessdebt}. To support this, card-based games help identify ethical~\cite{Alidoosti2023} and fairness~\cite{Zhang2023} issues in software design. Although these approaches help identify fairness issues, they do not directly guide the architectural design process. Our DT framework supplements these approaches by embedding fairness considerations through an exploratory SA-focused DT approach. 

Architectural choices can be used to mitigate bias, for example, by adopting fairness-aware tactics~\cite{Yi2022} or by introducing design patterns that enable fair algorithmic strategies~\cite{Vora2025}. 
To operationalize fairness at design time, our framework incorporates architectural tactics that translate abstract fairness concerns into concrete design choices. Complementing it, we use KPIs, which can help monitor the impact of fairness-supporting tactics. As important as including fairness at design time, it is equally important to measure it in terms of its impacts over time to understand how well the chosen fairness strategies work and change them based on the knowledge acquired over time~\cite{Castelnovo2021FairnessTime}, which is supported by our DT approach.  

\section{Threats to Validity}
\label{sec:threats}
In this section, we present possible threats to the validity and related mitigations. Threats are categorized according to the checklist by Peterson~\etal~\cite{PetersenK2013WRMa} for qualitative studies. 

\textbf{\textit{Construct Validity.}}
In this study, we only considered three foundational fairness types relevant to the use cases and the scope of the study. We do not use the concepts of retributive or informational fairness. The former deals with penalizing people for unethical behavior, e.g., on the moderation of social networks. The latter deals with issues such as information bias. Both fall outside the scope of SA (in data collection or algorithm implementation). Although our focus was on fairness, we used sustainability-specific frameworks in problem and solution spaces. The literature shows that both are multidimensional concepts, where fairness is sometimes studied under the umbrella of sustainability~\cite{ponsard2024}. We only used these frameworks to create a structured step-by-step process. The content used to initialize the frameworks was independent of the frameworks. 

\textbf{\textit{Internal Validity.}}
In thematic analysis, data coding and identification patterns are subject to bias. To mitigate this threat, the categorization is performed by multiple authors, and the changes are merged after a discussion process. As the study results are based on assignment data, it might raise questions about data quality. We mitigated this threat by setting an inclusion criterion for data selection.

\textbf{\textit{External Validity.}}
Due to a small sample size, the conclusions cannot be generalized for a specific domain or fairness type. However, the study provides a foundation for further research using this approach in larger and more refined data sets, e.g., in the industrial domain. Although the DT approach is used for fairness, it can be extended to similar multidimensional quality concerns, such as sustainability, fairness, and trustworthiness, etc.

\textbf{\textit{Reliability.}}
For transparency, we provide data from all three phases in our replication package. 
All thematic analysis steps, such as coding schemes, categorization, and identification of patterns, were cross-checked by all authors. In case of disagreements, a consensus was reached after discussion.

\section{Conclusion and Future Work}
\label{sec:conclusion}
In this study, we presented a fairness-first DT approach for SA design, executed in an educational context. Our results show that (i) context is key to solving fairness issues, (ii) multiple fairness types must be considered during problem elaboration, (iii) composite views with contextual elements can aid in preservation of contextual details and decision making logic, (vi) fairness perspective helps define scope of fairness, provides a roadmap for monitoring it via KPIs, and makes it actionable through tactics, (v) design views provide a visualization of end-to-end traceability between fairness quality concerns, architectural elements and operational context. As an outcome of this study, we also provide reusable templates and an education guide enabling adoption of our DT approach, made open source via our replication package. Our results and education guide are valuable for educators seeking to integrate socio-technical concerns into SA education.
Future work involves integrating the learnings of this study into the course design, collecting student feedback, and understanding the needs and processes of the industry to find entry points to incorporate fairness-first DT. 

\begin{acks}
We are grateful to the students of the Fair Digital Design for consenting to the analysis of their assignment data for this research.
This publication is part of the project SustainableCloud (OCENW.M20.243) of the research programme Open Competition by the Dutch Research Council (NWO).
We acknowledge the use of the AI tool \textit{Writeful} tool for grammatical improvements in the text of this paper.
\end{acks}
\bibliographystyle{ACM-Reference-Format}
\bibliography{references}

@inproceedings{FunkeLago_Perspectives_2025,
	title        = {Towards an {{Architectural Perspective}} for {{Sustainability}}: {{Bundle}} the {{Needs}} from {{Industry}}},
	author       = {Funke, Markus and Lago, Patricia},
	year         = 2025,
	booktitle    = {Quality of Information and Communications Technology},
	publisher    = {Springer Nature Switzerland},
	note         = {In press \url{https://arxiv.org/abs/2508.20774}}
}

@article{Lago2024-SAF-Toolkit,
	title        = {The sustainability assessment framework toolkit: a decade of modeling experience: The sustainability assessment framework toolkit: a decade of modeling experience},
	author       = {Lago, Patricia and Condori-Fernandez, Nelly and Fatima, Iffat and Funke, Markus and Malavolta, Ivano},
	year         = 2024,
	journal      = {Softw. Syst. Model.},
	publisher    = {Springer-Verlag},
	volume       = 24,
	number       = 2,
	doi          = {10.1007/s10270-024-01230-9}
}

@article{KPI-Fatima-2024,
	title        = {Providing Guidance to Software Practitioners: A Framework for Creating KPIs},
	author       = {Fatima, Iffat and Funke, Markus and Lago, Patricia},
	year         = 2024,
	journal      = {IEEE Software},
	volume       = {},
	number       = {},
	doi          = {10.1109/MS.2024.3456446}
}

@misc{santos2025softwarefairnessdebt,
	title        = {Software Fairness Debt},
	author       = {Ronnie de Souza Santos and Felipe Fronchetti and Savio Freire and Rodrigo Spinola},
	year         = 2025,
	url          = {https://arxiv.org/abs/2405.02490},
	eprint       = {2405.02490},
	archiveprefix = {arXiv},
	primaryclass = {cs.SE}
}

@article{deSouzaSantos2025--biassmells,
	title        = {Bias Smells: Exploring the Roots of Discrimination and Societal Inequities in Software Systems},
	author       = {de Souza Santos, Ronnie and Magalhaes, Cleyton and Spínola, Rodrigo},
	year         = 2025,
	journal      = {SSRN Electronic Journal},
	doi          = {10.2139/ssrn.5227890}
}

@article{Obermeyer2019-bias-health,
	title        = {Dissecting racial bias in an algorithm used to manage the health of populations},
	author       = {Ziad Obermeyer  and Brian Powers  and Christine Vogeli  and Sendhil Mullainathan},
	year         = 2019,
	journal      = {Science},
	volume       = 366,
	number       = 6464,
	doi          = {10.1126/science.aax2342}
}

@article{Hurlin2024-creditscoring,
	title        = {The Fairness of Credit Scoring Models},
	author       = {Hurlin, Christophe and P\'{e}rignon, Christophe and Saurin, S\'{e}bastien},
	year         = {2024},
	journal      = {Management Science},
	doi          = {10.1287/mnsc.2022.03888}
}

@inproceedings{Abrams2025,
    author="Abrams, Kyra Milan",
    title="Digital Redlining: Past and Present Motivations",
    booktitle="Advances in Information and Communication",
    year="2025",
    publisher="Springer Nature Switzerland",
    address="Cham",
    pages="32--43",
    isbn="978-3-031-85363-0"
}

@book{jefferson2020digitize,
	title        = {Digitize and Punish: Racial Criminalization in the Digital Age},
	author       = {Jefferson, Brian},
	year         = 2020,
	publisher    = {University of Minnesota Press},
	isbn         = 9781517909222
}

@article{Kelly-Lyth2021,
	title        = {Challenging Biased Hiring Algorithms},
	author       = {Kelly-Lyth, Aislinn},
	year         = 2021,
	journal      = {Oxford Journal of Legal Studies},
	volume       = 41,
	number       = 4,
	doi          = {10.1093/ojls/gqab006}
}

@article{Golbeck2025,
	title        = {Recommender System-Induced Eating Disorder Relapse: Harmful Content and the Challenges of Responsible Recommendation},
	author       = {Golbeck, Jennifer},
	year         = 2025,
	journal      = {ACM Trans. Intell. Syst. Technol.},
	publisher    = {Association for Computing Machinery},
	volume       = 16,
	number       = 1,
	doi          = {10.1145/3675404}
}

@article{Ryan2023,
	title        = {Integrating Fairness in the Software Design Process: An Interview Study With HCI and ML Experts},
	author       = {Ryan, Seamus and Nadal, Camille and Doherty, Gavin},
	year         = 2023,
	journal      = {IEEE Access},
	volume       = 11,
	number       = {},
	doi          = {10.1109/ACCESS.2023.3260639}
}

@inproceedings{Alidoosti2022,
	title        = {Incorporating Ethical Values into Software Architecture Design Practices},
	author       = {Alidoosti, Razieh and Lago, Patricia and Poort, Eltjo and Razavian, Maryam and Tang, Antony},
	year         = 2022,
	booktitle    = {2022 IEEE 19th International Conference on Software Architecture Companion (ICSA-C)},
	volume       = {},
	number       = {},
	doi          = {10.1109/ICSA-C54293.2022.00031}
}

@article{Zhang2023,
	title        = {Fairness in Design: A Framework for Facilitating Ethical Artificial Intelligence Designs},
	author       = {Zhang, Jiehuang and Shu, Ying and Yu, Han},
	year         = 2023,
	journal      = {International Journal of Crowd Science},
	volume       = 7,
	number       = 1,
	doi          = {10.26599/IJCS.2022.9100033}
}

@inproceedings{Alidoosti2023,
	title        = {Designing Ethics-Aware DecidArch Game to Promote Value Diversity in Software Architecture Design Decision Making},
	author       = {Alidoosti, Razieh and Lago, Patricia and Poort, Eltjo and Razavian, Maryam},
	year         = 2023,
	booktitle    = {Universal Access in Human-Computer Interaction},
    publisher="Springer Nature Switzerland",
    address="Cham",
    pages="3--26",
    
}

@inproceedings{Yi2022,
	title        = {The larger the fairer? small neural networks can achieve fairness for edge devices},
	author       = {Sheng, Yi and Yang, Junhuan and Wu, Yawen and Mao, Kevin and Shi, Yiyu and Hu, Jingtong and Jiang, Weiwen and Yang, Lei},
	year         = 2022,
	booktitle    = {Proceedings of the 59th ACM/IEEE Design Automation Conference},
	publisher    = {Association for Computing Machinery},
	doi          = {10.1145/3489517.3530427},
	isbn         = 9781450391429
}

@article{Shelley2012Fairness,
	title        = {Fairness in Technological Design},
	author       = {Cameron Shelley},
	year         = 2012,
	journal      = {Science and Engineering Ethics},
	publisher    = {Springer},
	volume       = 18,
	number       = 4,
	doi          = {10.1007/s11948-011-9259-1}
}

@inproceedings{Castelnovo2021FairnessTime,
	title        = {Towards Fairness Through Time},
	author       = {Alessandro Castelnovo and Luca Malandri and Federico Mercorio and Mario Mezzanzanica and Alberto Cosentini},
	year         = 2021,
	booktitle    = {Machine Learning and Principles and Practice of Knowledge Discovery in Databases},
	publisher    = {Springer, Cham},
	volume       = 1524,
	doi          = {10.1007/978-3-030-93736-2_46}
}

@inproceedings{Selbst2019,
	title        = {Fairness and Abstraction in Sociotechnical Systems},
	author       = {Selbst, Andrew D. and Boyd, Danah and Friedler, Sorelle A. and Venkatasubramanian, Suresh and Vertesi, Janet},
	year         = 2019,
	booktitle    = {Proceedings of the Conference on Fairness, Accountability, and Transparency},
	publisher    = {Association for Computing Machinery},
	doi          = {10.1145/3287560.3287598},
	isbn         = 9781450361255
}

@book{bass2012software,
	title        = {Software Architecture in Practice},
	author       = {Len Bass and Paul Clements and Rick Kazman},
	year         = 2012,
	publisher    = {Addison-Wesley},
	isbn         = 9780321815736,
	edition      = {3rd}
}

@incollection{koh2015designthinking,
	title        = {Design Thinking and Education},
	author       = {Koh, Joyce H. L. and Chai, Ching Sing and Wong, Boon and Hong, Huang-Yao},
	year         = 2015,
	booktitle    = {Design Thinking for Education},
	publisher    = {Springer, Singapore},
	doi          = {10.1007/978-981-287-444-3_1}
}

@article{steen2013codesign,
	title        = {Co-Design as a Process of Joint Inquiry and Imagination},
	author       = {Marc Steen},
	year         = 2013,
	journal      = {Design Issues},
	volume       = 29,
	number       = 2,
	doi          = {10.1162/DESI_a_00207}
}

@inproceedings{Valentim2017,
	title        = {The Students' Perspectives on Applying Design Thinking for the Design of Mobile Applications},
	author       = {Costa Valentim, Natasha M. and Silva, Williamson and Conte, Tayana},
	year         = 2017,
	booktitle    = {2017 IEEE/ACM 39th International Conference on Software Engineering: Software Engineering Education and Training Track (ICSE-SEET)},
	volume       = {},
	number       = {},
	doi          = {10.1109/ICSE-SEET.2017.10}
}

@book{plattner2010designthinking,
	title        = {Design Thinking: Understand – Improve – Apply},
	author       = {Hasso Plattner and Christoph Meinel and Larry Leifer},
	year         = 2010,
	publisher    = {Springer},
	isbn         = 9783642137570
}

@inproceedings{Palacin-Silva2017,
	title        = {Infusing Design Thinking into a Software Engineering Capstone Course},
	author       = {Palacin-Silva, Maria and Khakurel, Jayden and Happonen, Ari and Hynninen, Timo and Porras, Jari},
	year         = 2017,
	booktitle    = {2017 IEEE 30th Conference on Software Engineering Education and Training (CSEE\&T)},
	volume       = {},
	number       = {},
	doi          = {10.1109/CSEET.2017.41}
}

@inproceedings{Cleland-Huang2017,
	title        = {A case study: Injecting safety-critical thinking into graduate software engineering projects},
	author       = {Cleland-Huang, Jane and Rahimi, Mona},
	year         = 2017,
	booktitle    = {Proceedings of the 39th International Conference on Software Engineering: Software Engineering and Education Track},
	publisher    = {IEEE Press},
	doi          = {10.1109/ICSE-SEET.2017.4},
	isbn         = 9781538626719
}

@inproceedings{Gama2018,
	title        = {Combining Challenge-Based Learning and Design Thinking to Teach Mobile App Development},
	author       = {Gama, Kiev and Castor, Fernando and Alessio, Pedro and Neves, Andre and Araújo, Cristiano and Formiga, Rafael and Soares-Neto, Francisco and Oliveira, Higor},
	year         = 2018,
	booktitle    = {2018 IEEE Frontiers in Education Conference (FIE)},
	volume       = {},
	number       = {},
	doi          = {10.1109/FIE.2018.8658447}
}

@book{aristotle2014nicomachean,
	title        = {The Nicomachean Ethics},
	author       = {Aristotle},
	year         = 2014,
	publisher    = {Penguin Classics},
}

@article{leventhal1980procedural,
	title        = {What Should Be Done with Equity Theory?},
	author       = {Gerald S. Leventhal},
	year         = 1980,
	journal      = {Social Exchange: Advances in Theory and Research}
}

@inproceedings{bies1986interactional,
	title        = {Interactional Justice: Communication Criteria of Fairness},
	author       = {Robert J. Bies and Joseph S. Moag},
	year         = 1986,
	booktitle    = {Annual Meeting of the Academy of Management},
    publisher    =     {JAI Press},
}

@article{greenberg1993justice,
	title        = {The Social Side of Fairness: Interpersonal and Informational Classes of Organizational Justice},
	author       = {Jerald Greenberg},
	year         = 1993,
	journal      = {Justice in the Workplace},
	volume       = 1
}

@article{carlsmith2002psychological,
	title        = {Why Do We Punish? Deterrence and Just Deserts as Motives for Punishment},
	author       = {Kevin M. Carlsmith and John M. Darley and Paul H. Robinson},
	year         = 2002,
	journal      = {Journal of Personality and Social Psychology},
	volume       = 83,
	number       = 2, 
}

@inproceedings{dwork2012fairness,
	title        = {Fairness Through Awareness},
	author       = {Cynthia Dwork and Moritz Hardt and Toniann Pitassi and Omer Reingold and Richard Zemel},
	year         = 2012,
	booktitle    = {Proceedings of the 3rd Innovations in Theoretical Computer Science Conference (ITCS)},
	publisher    = {ACM},
	doi          = {10.1145/2090236.2090255}
}

@article{kleinberg2016inherent,
	title        = {Inherent Trade-Offs in the Fair Determination of Risk Scores},
	author       = {Jon Kleinberg and Sendhil Mullainathan and Manish Raghavan},
	year         = 2016,
	journal      = {Proceedings of Innovations in Theoretical Computer Science (ITCS)}
}

@article{corbett2018measure,
    author = {Corbett-Davies, Sam and Gaebler, Johann D. and Nilforoshan, Hamed and Shroff, Ravi and Goel, Sharad},
    title = {The measure and mismeasure of fairness},
    year = {2023},
    issue_date = {January 2023},
    publisher = {JMLR.org},
    volume = {24},
    number = {1},
    issn = {1532-4435},
    articleno = {312},
    numpages = {117},
}

@book{simon1969,
	title        = {The Sciences of the Artificial},
	author       = {Herbert A. Simon},
	year         = 1969,
	publisher    = {MIT Press}
}

@book{rawls1971theory,
	title        = {A Theory of Justice},
	author       = {John Rawls},
	year         = 1971,
	publisher    = {Harvard University Press}
}

@dataset{rep-pkg,
author       = {Iffat Fatima and
                  Funke, Markus and
                  Lago, Patricia},
  title        = {Replication Package: Fairness-First Design
                   Thinking for Software Architecture (Version 1)
                  },
  year         = 2025,
  publisher    = {Zenodo},
  version      = {1.0},
  doi          = {10.5281/zenodo.17977756},
  url          = {https://doi.org/10.5281/zenodo.17977756},
}

@book{rozanski2012,
	title        = {Software Systems Architecture: Working With Stakeholders Using Viewpoints and Perspectives},
	author       = {Rozanski, Nick and Woods, Eoin},
	year         = 2012,
	publisher    = {Addison-Wesley},
	edition      = {2nd}
}

@misc{iso42010,
	title        = {{ISO/IEC/IEEE 42010:2011 Systems and software engineering — Architecture description}},
	year         = 2011,
	note         = {Available at: \url{https://www.iso.org/standard/50508.html}},
	author = {{International Organization for Standardization}},
	institution  = {ISO/IEC/IEEE}
}

@book{hofmeister2007,
	title        = {Applied Software Architecture},
	author       = {Hofmeister, Christine and Nord, Robert L. and Soni, Dilip},
	year         = 2007,
	publisher    = {Addison-Wesley Professional}
}

@article{colquitt2001,
	title        = {On the dimensionality of organizational justice: A construct validation of a measure},
	author       = {Colquitt, Jason A.},
	year         = 2001,
	journal      = {Journal of Applied Psychology},
	publisher    = {American Psychological Association},
	volume       = 86,
	number       = 3,
	doi          = {10.1037/0021-9010.86.3.386}
}

@inproceedings{Migliorini2024,
	title={Architectural Views: The State of Practice in Open-Source Software Projects},
  author={Migliorini, Sofia and Verdecchia, Roberto and Malavolta, Ivano and Lago, Patricia and Vicario, Enrico},
  booktitle={European Conference on Software Architecture},
  pages={396--415},
  year={2024},
  organization={Springer}
}

@article{hsieh2005,
	title        = {Three Approaches to Qualitative Content Analysis},
	author       = {Hsieh, Hsiu-Fang and Shannon, Sarah E.},
	year         = 2005,
	journal      = {Qualitative Health Research},
	volume       = 15,
	number       = 9,
	doi          = {10.1177/1049732305276687}
}

@book{strauss1990open,
	title        = {Basics of Qualitative Research: Grounded Theory Procedures and Techniques},
	author       = {Strauss, Anselm and Corbin, Juliet},
	year         = 1990,
	publisher    = {Sage Publications}
}

@article{bingham2022apriori,
	title        = {From Data Management to Actionable Findings: A Five-Phase Process of Qualitative Data Analysis},
	author       = {Andrea J. Bingham},
	year         = 2022,
	journal      = {Sage Open},
	volume       = 12,
	number       = 2,
}

@inproceedings{PetersenK2013WRMa,
	 author={Petersen, Kai and Gencel, Cigdem},
  booktitle={2013 Joint Conference of the 23rd International Workshop on Software Measurement and the 8th International Conference on Software Process and Product Measurement}, 
  title={Worldviews, Research Methods, and their Relationship to Validity in Empirical Software Engineering Research}, 
  year={2013},
  volume={},
  number={},
  pages={81-89},
  doi={10.1109/IWSM-Mensura.2013.22}
}

@misc{ponsard2024,
	title        = {Modelling and Classification of Fairness Patterns for Designing Sustainable Information Systems},
	author       = {Christophe Ponsard and Bérengère Nihoul and Mounir Touzani},
	year         = 2024,
	url          = {https://arxiv.org/abs/2411.17894},
	eprint       = {2411.17894},
	archiveprefix = {arXiv},
	primaryclass = {cs.SE}
}

@inproceedings{Vora2025,
	title        = {Architecting for Fairness},
	author       = {Vora, Urjaswala},
	year         = 2025,
	booktitle    = {2025 IEEE 22nd International Conference on Software Architecture Companion (ICSA-C)},
	volume       = {},
	number       = {},
	doi          = {10.1109/ICSA-C65153.2025.00052}
}

@article{Lago2015-sus-quality,
	title        = {Framing sustainability as a property of software quality},
	author       = {Lago, Patricia and Ko\c{c}ak, Sedef Akinli and Crnkovic, Ivica and Penzenstadler, Birgit},
	year         = 2015,
	journal      = {Commun. ACM},
	publisher    = {ACM},
	volume       = 58,
	number       = 10,
	doi          = {10.1145/2714560},
	url          = {https://doi.org/10.1145/2714560}
}

@inproceedings{Jansen2005,
	title        = {Software Architecture as a Set of Architectural Design Decisions},
	author       = {Jansen, A. and Bosch, J.},
	year         = 2005,
	booktitle    = {5th Working IEEE/IFIP Conference on Software Architecture (WICSA'05)},
	volume       = {},
	number       = {},
	doi          = {10.1109/WICSA.2005.61},
}

@techreport{dSchool2009,
  title        = {An Introduction to Design Thinking Process Guide},
  author       = {{Hasso Plattner Institute of Design at Stanford}},
  institution  = {Stanford University},
  year         = {2009},
  address      = {Stanford, CA, USA},
  type         = {Technical Report},
}
\end{document}